\documentclass[twocolumn]{aa}
\usepackage{graphicx}
\usepackage{natbib}
\begin{document}
 \title{VLT narrow-band photometry in the Lyman continuum of two galaxies at
 $z\sim3$\thanks{Based on observations collected at the European Southern
 Observatory, Paranal (Chile); Proposal No.: 71.A-0418(A,B).}}

 \subtitle{Limits to the escape of ionizing flux}

   \author{A. K. Inoue\thanks{Postdoctoral Fellow of the JSPS for Research Abroad.}
          \inst{1,2,3}
          \and
          I. Iwata\inst{3,4}\thanks{Present address: Okamaya Astrophysical
          Observatory, Honsho, Kamogata-cho, Asakuchi-gun, Okayama 719-0232,
          Japan}
          \and
          J.-M. Deharveng\inst{1}
          \and
          V. Buat\inst{1}
          \and
          D. Burgarella\inst{1}
          }

   \offprints{A. K. Inoue}

   \institute{Laboratoire d'Astrophysique de Marseille, Traverse du
              Siphon, BP 8, 13376 Marseille, CEDEX 12, France\\
              \email{akio.inoue@oamp.fr,veronique.buat@oamp.fr,denis.burgarella@oamp.fr,jean-michel.deharveng@oamp.fr}
              \and
              Department of Physics, Kyoto University, 
              Sakyo-ku, Kyoto 606-8502, Japan
              \and
              Department of Astronomy, Kyoto University, 
              Sakyo-ku, Kyoto 606-8502, Japan
             \and
              Subaru Mitaka Office (Subaru Telescope), 
              National Astronomical Observatory of Japan, 
              2-21-1 Osawa, Mitaka, Tokyo 181-8588, Japan\\
              \email{iwata@optik.mtk.nao.ac.jp}
             }

   \date{Submitted on August 2, 2004}

   \abstract{

   We have performed narrow-band imaging observations with the 
   Very Large Telescope, aimed at detecting the Lyman continuum (LC) 
   flux escaping from galaxies at $z\sim3$. We do not find any significant 
   LC flux from our sample of two galaxies in the Hubble Deep Field South, 
   at $z=3.170$ and $3.275$. The corresponding lower limits on the 
   $F_{1400}/F_{900}$ flux density (per Hz) ratio are 15.6 and 10.2 
   (3-$\sigma$ confidence level). After correction for the intergalactic
   hydrogen absorption, the resulting limits on the relative escape fraction
   of the LC are compared with those obtained by different approaches, at
   similar or lower redshifts. One of our two objects has a relative escape
   fraction lower than the detection reported by Steidel et al.\ in a
   composite spectrum of $z\sim3$ galaxies. A larger number of objects is
   required to reach a significant conclusion. Our comparison shows the
   potential of narrow-band imaging for obtaining the best limit on the
   relative escape fraction at $z\sim3$. Stacking a significant number
   of galaxies observed through a narrow-band filter would provide constraint
   on the galactic contribution to the cosmic reionization.

   }

   \titlerunning{Upper limit of Lyman continuum escape}
   \authorrunning{Inoue et al.}

   \maketitle
%
\section{Introduction}

The observations of the finishing and the beginning reionization of
intergalactic hydrogen at $z \sim 6$ and 20 \citep{bec01,kog03} 
combined with the later reionization of intergalactic He {\sc ii} 
\citep{kri01,zhe04} at $z\sim 3$ are now placing constraints on the
nature and the evolution of the ultra-violet (UV) background radiation.
A dominant contribution of star-forming galaxies is suggested at $z>3$
before the density of luminous quasars become significant. This picture
would gain additional support from direct measurements of the amount of
Lyman continuum (LC) radiation released by galaxies into the
intergalactic medium (IGM). 

Direct measurements of the LC radiation from galaxies are, however, very 
challenging. In the local universe, they require space instruments 
sensitive at short UV wavelengths as well as galaxies slightly redshifted 
to avoid the neutral hydrogen opacity (continuum and line absorption) 
in the Galaxy. 
Observations of nearby star-forming galaxies with the Hopkins Ultraviolet
Telescope (HUT) and the {\it Far Ultraviolet Spectroscopic Explorer (FUSE)}
have led so far only to upper limits on the fraction of LC (or emitted 900
\AA) photons that escape; the so-called escape fraction ($f_{\rm esc}$) is 
typically less than about 10\% \citep{lei95,hur97,deh01}.
This is consistent with the values derived from the near-blackness of 
interstellar lines tracing the H {\sc i} gas as observed with {\it FUSE} 
in five other starburst galaxies \citep{hec01}.

At high redshifts, the increasing IGM opacity (absorption of the
ionizing radiation by the neutral hydrogen), not to speak of galaxies 
becoming fainter, makes the observations difficult.
Nevertheless, large ground-based telescopes can enter the competition at
$z\ga3$. \cite{ste01} (hereafter S01) derived a 1500\AA/900\AA\ observed 
flux density\footnote{In this paper, all flux and luminosity densities 
are given per unit frequency interval.} ratio, 
$(F_{1500}/F_{900})_{\rm obs}=17.7\pm3.8$ from a
composite spectrum of 29 Lyman break galaxies (LBGs) at a mean redshift of 
3.4 with Keck/LRIS. By comparison with models of the UV spectral energy
distribution of star-forming galaxies, the flux density ratio, corrected for
the IGM opacity, leads to the fraction of escaping LC (900\AA) photons
relative to the fraction of escaping non-ionizing UV (1500\AA) photons. 
This is called the {\it relative} escape fraction ($f_{\rm esc,rel}$) 
and is different from the first definition of the escape fraction 
($f_{\rm esc}$) used above for nearby galaxies. S01 interpreted  
their results as implying $f_{\rm esc,rel}\ga50\%$.

However, not all $z\sim3$ galaxies emit a significant LC radiation. 
Using the FORS2 spectrograph on the VLT, \cite{gia02} (hereafter G02) 
obtained 1-$\sigma$ lower limits on $(F_{1500}/F_{900})_{\rm obs}$ four times
larger than the value of S01 for two bright LBGs. \cite{hec01}
also deduced a very low escape fraction for a gravitational lensed galaxies,
MS 1512-cB58 at $z=2.7$ from the detailed analysis of interstellar
absorption lines. 

Observations have not been only spectroscopic. Imaging of galaxies
at $z\sim1$ with the FUV solar-blind detector of Space Telescope
Imaging Spectrograph (STIS) have also provided constraints on the 
flux below the Lyman limit \citep{fer01,mal03}. In particular, 
\cite{mal03} (hereafter M03) have obtained lower limits of 
$(F_{1500}/F_{700})_{\rm obs} \ga 150$--1000 (1-$\sigma$),  
implying much lower LC escape fraction than in the galaxies of S01.
\cite{fer03} (hereafter FS03) have used the deep $U_{300}$ images of the
Hubble Deep Field North (HDFN) and reported an average LC escape fraction
of no more than 4\% for 27 galaxies at redshifts $1.9 < z < 3.5$. 
As a number of galaxies in their sample have their $U_{300}$ fluxes
contaminated by nonionizing UV photons, their analysis is based on models and
indirect. 

These conflicting results, as well as the difficulties of observations 
at high redshift, have led us to search for possibilities of improvement.
We have first examined how the detection advantage of imaging over
spectroscopy is actually working in the specific context of measuring faint LC
radiation.
On one hand, spectroscopy allows us to measure the LC flux close to the
redshifted Lyman limit, where the average IGM opacity is not yet as large as
it would be at shorter wavelength because of the Lyman valley 
\citep[e.g.,][]{mol90}.  
On the other hand, broad-band measurements require a very large correction for
the IGM opacity as we show in the current paper, because the effective
wavelength becomes very short for objects selected at a redshift appropriate
for avoiding contamination by nonionizing UV photons. The narrow-band imaging
appears, therefore, as a natural compromise between the lower opacity offered
by spectroscopy and the depth of detection offered by broad-band imaging.

We have, for the first time, attempted imaging observations from the
ground, using the VLT, aimed at measuring the LC radiation from galaxies at
high redshifts through a narrow-band filter. The goal of this paper is 
twofold: 
(i) reporting the data and their interpretation in terms of the relative 
escape fraction, and 
(ii) validating the narrow-band photometric approach by quantitative 
comparisons. 

In Sect.~2, we describe the details of our strategy, the choice 
of the filter, the target selection based on photometric redshifts, 
the spectroscopic observations for confirming redshifts and the imaging 
observations. Data reduction and results are described in Sect.~3. 
Their interpretation in terms of {\it relative} escape fraction are presented 
in Sect.~4. A comparison of the constraints reached with those from 
other observations is given in Sect.~5, where we also discuss the 
{\it absolute} escape fraction. The final section is devoted 
to our conclusions.

We adopt a flat $\Lambda$-dominated cosmology with $\Omega_\Lambda=0.7$ and
$\Omega_{\rm m}=0.3$, and the current Hubble constant $H_0=70$ km
s$^{-1}$ Mpc$^{-1}$ to estimate luminosities.

\section{Observations}

\subsection{Filter selection}

A direct and unambiguous detection (or upper limit) of the LC escaping 
from galaxies requires a lower redshift limit of the target galaxies to push
out all emission red-ward of the Lyman limit beyond the long wavelength
cut-off of the filter transmission. The cut-off wavelength is
defined practically at 1\% transmission relative to the peak.
Since the IGM opacity rapidly increases with the source redshift
(e.g., see Fig.~A.2), the limiting redshift should be set as low as possible. 
The $OII+44$ narrow-band filter, whose long wavelength cut-off is the
smallest among all filters of FORS on VLT, was selected for our observation.
Accordingly, galaxies with $z\geq3.18$ must be selected.
We note here that the long wavelength cut-off of this filter 
is smaller than that of the $U_{300}$ of WFPC2 on the HST as shown in Fig.~1.
Indeed, the redshift limit for the $U_{300}$ would be $z\geq3.33$.

\begin{figure}
 \centering
 \includegraphics[width=0.9\linewidth]{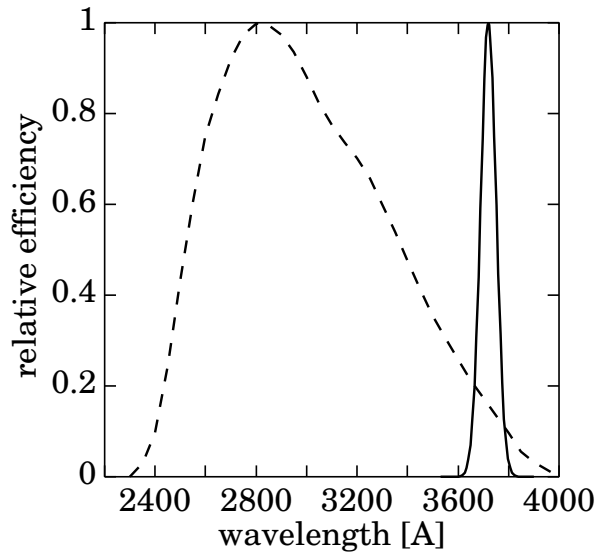}
 \caption{Transmission curves of the FORS $OII+44$ filter (solid) and
 the HST $U_{300}$ (dashed).}
\end{figure}

\subsection{Observed field and target galaxies}

As the redshift surveys in the southern celestial sphere had not produced 
a large number of galaxies with spectroscopic redshift ($z_{\rm sp}$) 
larger than 2, we had to select galaxies based on photometric redshift 
($z_{\rm ph}$) in the first phase of our observations. 
\cite{lab03} made a catalog of photometric redshift of galaxies 
in the Hubble Deep Field South (HDFS; \citealt{wil00}) by using their
ultra-deep near-infrared images obtained with VLT/ISAAC and HST/WFPC2 optical
images. Thus, we selected the HDFS/WFPC2 field as our target field. 

There are 24 galaxies with $I_{814}\leq25.0$ mag(AB)\footnote{The iso-photal 
magnitude in the HDFS WFPC2 catalog version 2 \citep{cas00}} and 
$z_{\rm ph}\geq2.5$ in the field. We set this lower limit of $z_{\rm ph}$, 
taking into account the 2-$\sigma$ uncertainty of $z_{\rm ph}$ 
for $z\sim3$ ($\sigma\approx 0.08(1+z)$ noted in \citealt{lab03}). 
Eight of them have $z_{\rm ph}$ larger than 3.18. On the other hand, 
six out of 24 have $z_{\rm sp}$, but all of them are less than 3.18.
We note that all the known $z_{\rm sp}$ are smaller than the corresponding
$z_{\rm ph}$.

An accurate redshift is essential to avoid contamination from nonionizing 
photons. We need a redshift accuracy of the order of 0.01.
Since the accuracy of $z_{\rm ph}$ is not sufficient, 
we performed spectroscopic observations with FORS2 on VLT using the MXU mode. 
A full description of the observation will be published elsewhere
(Iwata et al. in preparation). Here, we summarize the results briefly. 

Because of the limitation of the slit-let configuration, we were able to take
spectra of only 15 galaxies out of the above 24 galaxies. Five out of the 15
galaxies have known $z_{\rm sp}$ and 7 of them have $z_{\rm ph} > 3.18$.
The choice of galaxies was made with a priority given to more
luminous galaxies and galaxies with $z_{\rm ph} > 3.18$. Additionally, we took
spectra of galaxies with $z_{\rm ph}\geq2.5$ but $I_{814}>25.0$ mag(AB), 
for which we could configure the slit-let. Among these galaxies, one has 
$z_{\rm sp}$ previously measured. We ended up with obtaining 13 $z_{\rm sp}$: 
seven of them are new measurements and the remaining six are confirmations of
the previous results.  

Among the new redshifts, we got only one galaxy with a high enough redshift,
$z_{\rm sp}=3.275$ (HDFS 1825), in contrast to the expectation from photometric
redshifts. Indeed, for the 13 galaxies with $z_{\rm sp}$, 
we always find $z_{\rm sp} < z_{\rm ph}$, again. There is another galaxy with 
$z_{\rm sp}=3.170$ (HDFS 85) which was confirmed by our spectroscopy. 
This redshift is slightly smaller than the limiting redshift, however, 
the transmission efficiency of the $OII+44$ filter 
at the Lyman limit of the galaxy is as small as a few percent  
of the peak efficiency (see Fig.~1). The contamination of 
nonionizing photons is small enough, especially in the context of upper 
limit measurements. Therefore, our final sample consists of the two
galaxies. We have no other explanations for this small number of target
galaxies than a systematic error in the estimation of $z_{\rm ph}$.

\subsection{Imaging observations}

The HDFS field was observed through the $OII+44$ narrow-band filter
with  FORS1 and the TK2048EB4-1 detector chip in service mode from 29
June 2003 to 28 August 2003. The standard resolution collimator was
adopted, giving a pixel scale of $0.''2$ per pixel. The field of
view is $6.'8\times6.'8$ and covers the whole area of the HDFS/WFPC2
field. Each image was taken with a small dithering (typically $10''$)
and a typical exposure time of 1,080 sec.  The seeing size 
of each image was $0.''39$--$1.''23$, typically
$\sim0.''8$. We secured 38 images of the field.
The effective exposure time is 40,636 sec (i.e. about 11h).
Table 1 shows a summary of the observations.

\begin{table}
 \caption[]{Details of the imaging observations.}
 \setlength{\tabcolsep}{3pt}
 \footnotesize
 \begin{minipage}{\linewidth}
  \begin{tabular}{lc}
   \hline\hline
   Dates (year 2003) & June 29, 30; July 5, 28; August 1, 28\\
   Instrument & FORS1 on VLT\\
   Collimator & standard resolution\\
   CCD & $2048\times2046$ pixels\\
   Pixel scale & $0.''2$/pixel\\
   Field of view & $6.'8\times6.'8$\\
   Filter & $OII+44$\\
   Central wavelength & 371.7 nm\\
   Transmission FWHM & 7.3 nm\\
   Seeing & $\sim0.''8$\\
   Effective exposure time & 40,636 sec\\
   \hline
  \end{tabular}
 \end{minipage}
\end{table}%

\section{Data reduction and results}

\subsection{Final image}

\begin{figure*}
 \centering
 \includegraphics[width=\linewidth]{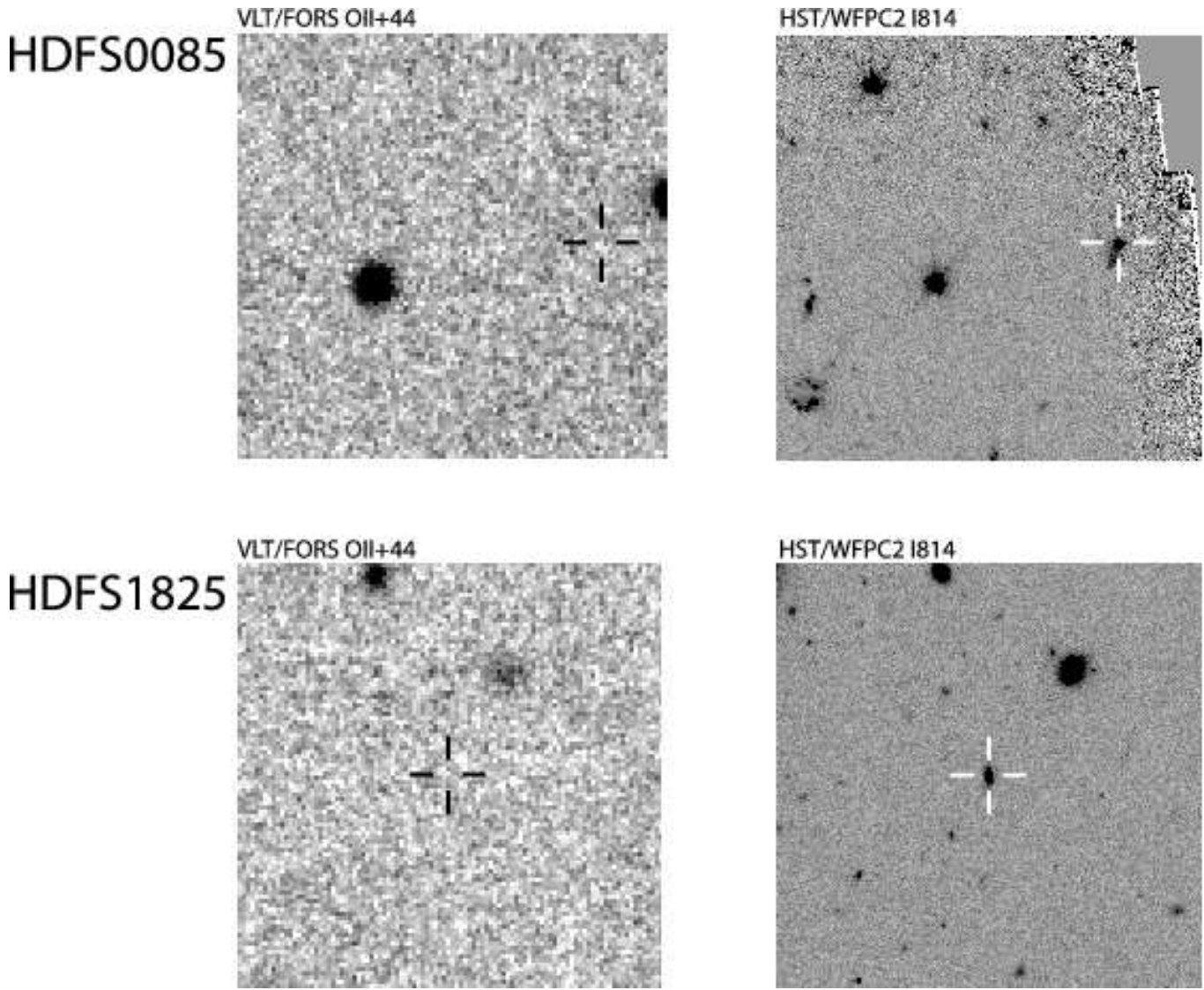}
 \caption{Close-up of sample galaxies through the $OII+44$ filter 
 (VLT/FORS, this work), and the $I_{814}$ filter \citep[HST/WFPC2,][]{wil00}. 
 The FOVs are $20'' \times 20''$. For the object 85, the image center is 
 $6''$ shifted from the object position, because it is close to 
 an edge of $I_{814}$ image.}
\end{figure*}

The image data reduction was carried out in a standard manner, 
using IRAF\footnote{IRAF is distributed by the National Optical 
Astronomy Observatories, which are operated by the Association of 
Universities for Research in Astronomy, Inc., under 
cooperative agreement with the National Science Foundation.}. 
The bias subtraction was made using over-scanned regions. 
Normalized twilight sky frames in each observing night were used 
for flat fielding. We selected 25 stars in the field and used them to
register the 38 frames. After registration, the rms of 
residual shifts was 0.03 pixel. These stars were also used for an 
airmass correction. We found a slope of magnitude dependence on 
airmass of $-0.45$, and we corrected observed counts as it would have 
been observed at the zenith. The stability of the observing conditions 
is confirmed by measuring the corrected counts of these stars; 
the rms errors in counts are less than 0.05 mag for 19 stars brighter than 22
mag(AB) through the $OII+44$ filter among the above 25 stars. 
Then, the IRAF task IMCOMBINE was used to sum the frames. We took 
averages of each pixel adopting a 3-$\sigma$ clipping. FWHM of 
stellar objects in the final image is $\sim 1''$. 
Fig.~2 shows the close-up images of the two sample galaxies through the
$OII+44$ filter and through the $I_{814}$ filter of HST/WFPC2.
Neither galaxy seems to be seen through the $OII+44$ filter.

\subsection{Photometry}

Three standard stars, Feige 110, G 93-48, and LTT 9491, were also observed
through the same filter. These frames were processed 
in the same way as the HDFS frames. The absolute AB magnitude of these 
stars were calculated from the spectral table provided on ESO web page. 
We found that the airmass dependence for the standard stars was $-0.58$ 
which is somewhat larger than that from stars in the HDFS frames. 
We derived the photometric zero point for the $OII+44$ filter at the zenith 
as 22.98 from the airmass slope of the standard stars. 
The zero point changes less than 0.02 mag if we adopt the airmass slope 
$-0.45$ which we used for correction of the HDFS frames.

To reduce background fluctuations, we apply a Gaussian smoothing 
to the final image adapted to the expected size of the sample galaxies through
the $OII+44$ filter. The size through the HST/WFPC2 $B_{450}$ filter is a
reasonable approximation because the central wavelength is close to that of
the $OII+44$ filter and we are seeing the light from massive stars in both
filters. Diameters of $0.''72$ (HDFS 85) and $0.''41$ (HDFS 1825) are
estimated from the half-light radii of the galaxies through the $B_{450}$
filter reported in the HDFS WFPC2 catalog version 2 \citep{cas00}. We
adopt $2.5 \times 2.5$ pix as the smoothing scale. The FWHM of the PSF in the
smoothed image is 8.33 pixel ($=1.''7$).

None of the sample galaxies was detected at a 3-$\sigma$ level in the smoothed 
$OII+44$ image. The upper limit on their flux density has been based on the
smoothed PSF size because they are smaller than the PSF. The measured rms
fluctuations within a square box of $1.''7 \times 1.''7$ in the smoothed image
around the sample galaxies translate into 3-$\sigma$ limiting magnitudes of
27.37 mag(AB) and 27.40 mag(AB) for HDFS 85 and 1825, respectively.

In Table 2, we summarize the photometric measurements of the two sample
galaxies. Photometric data with HST are taken from the HDFS WFPC2 catalog
version 2 \citep{cas00}. In this table, all upper limits are 3-$\sigma$ level. 
Our limiting magnitude is comparable with that of the HDFS in the $U_{300}$ for
the size of our sample galaxies, although the HDFS limiting magnitude for  
point source is deeper than ours \citep{cas00}.

\begin{table}
 \caption[]{Photometric properties of sample galaxies.}
 \setlength{\tabcolsep}{3pt}
 \footnotesize
 \begin{minipage}{\linewidth}
  \begin{tabular}{lcc}
   \hline\hline
   ID\footnote{HDFS WFPC2 catalog version 2 \citep{cas00}.} & 85 & 1825\\
   \hline
   $\alpha$ (J2000) & 22 32 46.91 & 22 32 52.03\\
   $\delta$ (J2000) & $-$60 31 46.9 & $-$60 33 42.6\\
   redshift & 3.170 & 3.275\\
   $OII+44$\footnote{3-$\sigma$ upper limits. $U_{300}$ values of HDFS 85 and
   1825 are reported in the HDFS WFPC2 catalog version 2. As these values are
   similar to the 1-$\sigma$ uncertainty (HDFS 85) or negative (HDFS
   1825), we have retained 3-$\sigma$ upper limits. Note that HDFS 85 is
   located near the edge of the WFPC2 image} (nJy) 
   & $<44.88$ & $<39.81$\\
   ${U_{300}}^b$ (nJy) & $<101.8$ & $<42.09$\\
   $B_{450}$\footnote{With 1-$\sigma$ errors} (nJy) & $392.08\pm12.02$ & 
   $181.98\pm6.51$\\
   ${V_{606}}^c$ (nJy) & $698.68\pm8.86$ & $404.91\pm4.26$\\
   ${I_{814}}^c$ (nJy) & $859.63\pm17.82$ & $504.49\pm7.58$\\ 
   \hline
  \end{tabular}
 \end{minipage}
\end{table}%

\section{Lyman continuum escape fraction}

From the upper limits obtained above on the LC flux of the two galaxies at 
$z=3.170$ and $z=3.275$, we try here to estimate the escape fraction of 
LC photons. As mentioned in the introduction, the term of escape fraction 
has been used in at least two ways. We first clarify the definition of 
the escape fractions.

\subsection{Definition of escape fractions}

We define two escape fractions. One is 
\begin{equation}
 f_{\rm esc} \equiv \frac{L_{\rm LC}^{\rm out}}{L_{\rm LC}^{\rm int}} 
  = \exp(-\tau_{\rm LC}^{\rm ISM})\,,
\end{equation}
where $L_{\rm LC}^{\rm int}$ is the intrinsic LC luminosity density (per Hz) 
of a galaxy, $L_{\rm LC}^{\rm out}$ is the LC luminosity density just outside
of the galaxy (not the observed one, see below), and $\tau_{\rm LC}^{\rm ISM}$
is the opacity of the interstellar medium (ISM) for LC photons in the galaxy. 
The other is  
\begin{equation}
 f_{\rm esc,rel} \equiv f_{\rm esc} 
  \left(\frac{L_{\rm UV}^{\rm out}}{L_{\rm UV}^{\rm int}}\right)^{-1}
  = f_{\rm esc} \exp(\tau_{\rm UV}^{\rm ISM})\,,
\end{equation}
where $L_{\rm UV}^{\rm int}$, $L_{\rm UV}^{\rm out}$, and 
$\tau_{\rm UV}^{\rm ISM}$ are the intrinsic luminosity density, outside
luminosity density, and the ISM opacity of the nonionizing UV photons,
respectively. Hereafter, the former is called the {\it absolute} escape 
fraction, and the latter is called the {\it relative} escape fraction.

The IGM opacity should be taken into account for photons with a wavelength
shorter than the Ly$\alpha$ line at the source rest-frame. Namely, 
the observed luminosity density becomes 
$L_\lambda^{\rm obs}=L_\lambda^{\rm out}\exp(-\tau_\lambda^{\rm IGM})$
for the rest-frame wavelength $\lambda<1216$ \AA, and 
$L_\lambda^{\rm obs}=L_\lambda^{\rm out}$ otherwise if the IGM dust
extinction is negligible. Indeed, based on observations of distant supernovae
and on the thermal history of the IGM, \cite{ino04} have recently shown that 
the IGM dust extinction against $z\sim3$ sources should be less than 1 mag 
in the observer's $B$-band.
Taking a wavelength longer than the Ly$\alpha$ line for the UV
wavelength (e.g., $\sim1500$ \AA), we have 
\begin{equation}
 f_{\rm esc,rel}=
  \frac{(L_{\rm UV}/L_{\rm LC})_{\rm int}}
  {(F_{\rm UV}/F_{\rm LC})_{\rm obs}}
  \exp(\tau_{\rm LC}^{\rm IGM})\,,
\end{equation}
where we have replaced the observed luminosity density ratio into the observed 
flux density ratio. Hence, the relative escape fraction can be estimated from
the observed UV-to-LC flux density ratio if we know the intrinsic
UV-to-LC luminosity density ratio and the IGM opacity. 
Moreover, the absolute escape fraction can be estimated from the relative 
escape fraction via equation (2) if we know the ISM opacity 
for nonionizing UV photons.

\subsection{Intrinsic luminosity density ratios}

The intrinsic UV-to-LC luminosity density ratio is still very uncertain
observationally. As noted by S01, {\it we must rely exclusively
on models}. Here we adopt the Starburst 99 model \citep{lei99}. From this 
model, we obtain $(L_{1500}/L_{900})_{\rm int}=1.5$--5.5 as shown in
Fig.~3, assuming a constant star formation rate, the Salpeter initial mass
function with the mass range of 0.1--100 $M_\odot$, and the metallicity of
0.001--0.02 (the solar value is 0.02). The luminosity density ratios mainly
depend on the duration since the onset of star formation; the ratio starts
from a small value, monotonically increases with time, and saturates 
at a larger value after several $10^8$ yrs. 
In this paper, we adopt $(L_{1500}/L_{900})_{\rm int}=3.0$ 
according to the value adopted by S01.\footnote{S01 adopted the value as 
the Lyman discontinuity, and assumed the flatness of the intrinsic spectrum 
between 912 and 1500 \AA.}

Since the UV and LC wavelengths actually depend on the filter adopted and the
source redshift, we have to adjust the intrinsic UV-to-LC luminosity density
ratios to the correct wavelengths. 
Some of the ratios, relative to 1500 \AA, are displayed in Fig.~3 as open
squares and summarized in Table~3. This table will be used to calculate the
UV-to-LC luminosity density ratio appropriate for each individual galaxy in
Sects.~4.4 and 5.1. As illustrated in Fig.~3, the uncertainties on the
luminosity density ratios are 20--50\%. 

\begin{table}
 \caption[]{Adopted intrinsic ratios of luminosity densities 
   relative to 1500 \AA.}
 \setlength{\tabcolsep}{3pt}
 \footnotesize
 \begin{minipage}{\linewidth}
  \begin{tabular}{lc}
   \hline\hline
   $x$ & $(L_{1500}/L_x)_{\rm int}$\footnote{The unit of luminosity densities  
   is erg s$^{-1}$ Hz$^{-1}$.}\\
   (\AA) & \\
   \hline
   700 & 4.0\\
   900 & 3.0\\
   1400 & 1.0\\
   1700 & 1.1\\
   2000 & 1.2\\
   2500 & 1.4\\
   \hline
  \end{tabular}
 \end{minipage}
\end{table}%

\begin{figure}
 \centering
 \includegraphics[width=0.9\linewidth]{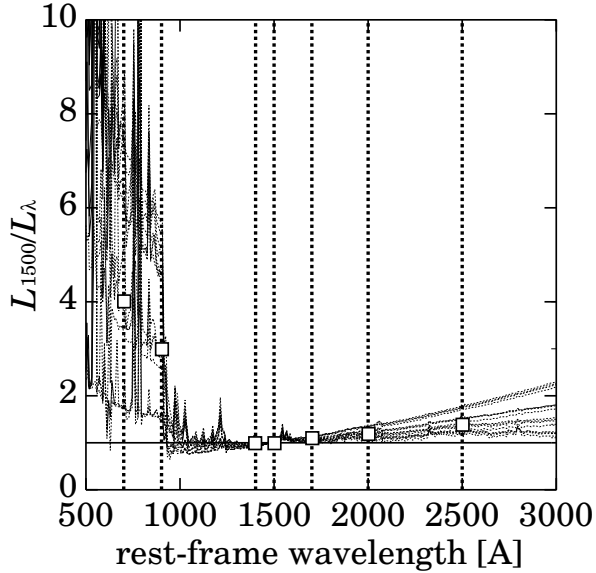}
 \caption{Intrinsic ratio of luminosity density at 1500 \AA\ to that at the 
 wavelength along horizontal axis. Thin dotted curves show 16 cases of the 
 combination of 4 metallicities (0.001, 0.004, 0.008, and 0.02) and 4
 ages (0 yr, 10 Myr, 100 Myr, and 1 Gyr), assuming a constant star
 formation and the Salpeter initial mass function with 0.1--100
 $M_\odot$. Open squares are fiducial ratios adopted in the text at 7
 wavelengths marked by thick dotted vertical lines. 
 Unit of luminosity densities is erg s$^{-1}$ Hz$^{-1}$.}
\end{figure}

\subsection{IGM opacity}

An average opacity of the IGM can be estimated from the number distribution
functions of the Lyman $\alpha$ forest and denser H {\sc i} absorbing clouds 
\citep[e.g.,][]{mad95}. The number of H {\sc i} clouds per unit interval of 
neutral hydrogen column density $N_{\rm HI}$ and per unit redshift $z$ 
interval can be expressed as 
$\partial^2 {\cal N}/\partial z \partial N_{\rm HI} \propto (1+z)^\gamma
N_{\rm HI}^{-\beta}$. \cite{kim02} have shown that the observed 
Lyman $\alpha$ forest data can be fitted with $\beta=1.5$, and 
$\gamma=2.5$ for $z\geq1.1$ and $\gamma=0.2$ for $z<1.1$, by combining their
high-$z$ data with the low-$z$ data of \cite{wey98}. 

In contrast to previous approaches \citep[e.g.,][]{mad95}, we have shown in
Appendix that not only the column density distribution but also the number
evolution of the denser clouds (essentially, Lyman limit systems) 
can be reasonably described by the same function as that obtained for
the Lyman $\alpha$ forest by \cite{kim02}. Based on 
this distribution function, we calculate an average IGM opacity,  
$\langle\tau^{\rm IGM}_{\lambda_{\rm obs}}(z_{\rm S})\rangle$, 
for a source with a redshift, $z_{\rm S}$, at an observed wavelength, 
$\lambda_{\rm obs}$ by an analytic approximation formula presented in Appendix
as equations (A.4--6).  
As seen in Appendix, our opacity is quantitatively very consistent 
with the value measured and adopted by S01, and shows a reasonable
agreement with the opacity model of \cite{ber99}. Finally, we integrate the IGM
opacity over the filter transmission when we make a correction for the IGM
absorption in order to derive the escape fraction from the photometric data. 
In the integration, we always assume a flat spectrum because we do not know 
the spectral shape of LC exactly.

\subsection{Estimation of relative escape fraction}

By combining the upper limit of the flux density through the $OII+44$
filter with the measurement of the flux density through the HST's $V_{606}$,
whose wavelength corresponds to the rest-frame $\sim 1400$ \AA\ for our
sample, we obtain a lower limit on the observed ratio 
$(F_{1400}/F_{900})_{\rm obs}$. This ratio has to be corrected for the
Galactic extinction before estimating the relative escape fraction.
For HDFS field, $E(B-V)=0.028$ mag \citep{sch98}, which corresponds to 
the Galactic extinction of $A_{V_{606}}- A_{OII}=-0.050$ mag if we adopt the
standard extinction law of our Galaxy \citep[e.g.,][]{whittet03}. 
Therefore, we obtain the 3-$\sigma$ lower limits of the corrected flux density
ratio $F_{1400}/F_{900}$ of 15.57 and 10.17 for HDFS 85 ($z=3.16$) and HDFS
1825 ($z=3.27$), respectively. 

Assuming $(L_{1400}/L_{900})_{\rm int}=3.0$ as found from Table 3 and  
mean IGM opacities integrated over the filter transmission, 
we obtain 3-$\sigma$ upper limits on the relative escape fraction as 
72\% and 216\% for HDFS 85 and 1825, respectively. As found in 
equation (2), the relative escape fraction is not restricted to 
less than 100\%, because it is the escape fraction {\it relative} to 
that of nonionizing UV photons. The obtained values are summarized 
in the top 2 lines of Table 4.

Since uncertain assumptions on the ISM dust extinction are inevitable for 
the calculation, the estimation of the absolute escape fraction is left
for the following section (\S 5.2), where we will discuss these issues in
detail.

\begin{table*}
 \caption[]{Summary of related observations and escape fractions.}
 \setlength{\tabcolsep}{3pt}
 \footnotesize
 \begin{minipage}{\linewidth}
  \begin{tabular}{lccccccccc}
   \hline\hline
   Redshift range/method/ & redshift & 
   $(F_{\rm UV}/F_{\rm LC})_{\rm obs}$\footnote{Significance
   level of lower/upper limits is 3-$\sigma$, but not included 
   uncertainties of model parameters. Uncertainty for Steidel's composite 
   spectrum means 1-$\sigma$ error of observations.} 
   & $\tau_{\rm LC}^{\rm IGM}$ & ${f_{\rm esc,rel}}^a$ & 
   $\lambda_{\rm UV},\lambda_{\rm LC}$\footnote{Wavelengths of ultraviolet
   (UV) and Lyman continuum (LC); 
     1: ($\lambda_{\rm UV},\lambda_{\rm LC})=(1400{\rm \AA},900{\rm \AA}$),  
     2: ($1500{\rm \AA},900{\rm \AA}$), 
     3: ($1400{\rm \AA},700{\rm \AA}$), 
     4: ($1700{\rm \AA},700{\rm \AA}$), 
     5: ($2000{\rm \AA},700{\rm \AA}$), 
     6: ($2500{\rm \AA},700{\rm \AA}$).}
   & $L_{\rm UV}$\footnote{UV luminosity density corrected for the Galactic
     extinction.}
   & $A_{\rm UV}$\footnote{In the estimation, $A_{1600}$--$\beta$ relation 
     by Meurer et al.(1999) is assumed, and $A_{1600}$ values are converted 
     to those at the appropriate UV wavelength by Calzetti's attenuation law 
     \citep{cal00}. For Steidel's composite spectrum, we have used the mean 
     $E(B-V)$ value reported in their paper.}
   & ref.\footnote{Reference of UV slope; 1: $V_{606}-I_{814}$ color in
     HST/WFPC2 catalog version 2 \citep{cas00}; 2: \cite{pet98}; 3: $B-V$
     color in M03; 4: $Un-G$ color in M03; 5: \cite{meu99}; 6: \cite{lei02}} 
   & ${f_{\rm esc}}^a$ \\ 
   galaxy name & & & & (\%) & & ($10^{29}$ erg s$^{-1}$ Hz$^{-1}$) & (mag) 
   &  & (\%)\\
   (1) & (2) & (3) & (4) & (5) & (6) & (7) & (8) & (9) & (10)\\
   \hline
   $z\sim3$ & & & & & & & & &\\
   \multicolumn{9}{l}{Narrow-band photometry (this work)} \\
   HDFS 85 & 3.170 & $>15.57$ & 1.32 & $<72$ & 1 & 1.59 & 1.6 & 1 
   & $<17$\\
   HDFS 1825 & 3.275 & $>10.17$ & 1.99 & $<216$ & 1 & 0.970 & 1.9 & 1 
   & $<38$\\
   Spectroscopy & & & & & & & &\\
   composite (29 galaxies) & 3.40 & $17.7\pm3.8$ & 1.50 & $76\pm16$ 
   & 2 & 1.60 & 0.26 & ... & $60\pm13$\\
   DSF 2237+116 C2 & 3.319 & $>24$ & 1.41 & $<51$ & 2 & 3.15 & 3.8 & 2 
   & $<1.5$\\
   Q0000-263 D6 & 2.961 & $>23$ & 1.04 & $<37$ & 2 & 4.85 & 2.3 & 2 & $<4.5$\\
   Broadband photometry & & & & & & & & &\\
   FLY99: 957 & 3.367 & $>2.34$ & 4.63 & ... & 3 & 0.116 & ... & ... & ...\\
   FLY99: 825 & 3.369 & $>11.9$ & 4.64 & ... & 3 & 0.617 & ... & ... & ...\\
   FLY99: 824 & 3.430 & $>7.89$ & 4.94 & ... & 3 & 0.638 & ... & ... & ...\\
   \hline
   $z\sim1$ & & & & & & & & &\\
   Broadband photometry & & & & & & & & &\\
   Cl J0023+0423:[LPO98a]022 & 1.1074 & $>187$ & 0.53 & $<3.0$ 
   & 5 & 0.967 & 1.3 & 3 & $<0.91$\\
   CFRS 03.1140 & 1.1818 & $>308$ & 0.59 & $<1.7$ & 6 & 2.56 & ... & ... 
   & $<1.7$\\
   CFRS 10.1887 & 1.2370 & $>119$ & 0.63 & $<4.5$ & 6 & 0.969 & ... & ... 
   & $<4.5$\\
   CFRS 10.0239 & 1.2919 & $>228$ & 0.68 & $<2.5$ & 6 & 2.28 & ... & ... 
   & $<2.5$\\
   CFRS 10.1168 & 1.1592 & $>163$ & 0.57 & $<3.1$ & 6 & 0.802 & ... & ... 
   & $<3.1$\\
   LDSS2 10.288 & 1.108 & $>102$ & 0.53 & $<5.5$ & 5 & 0.859 & ... & ... 
   & $<5.5$\\
   HDF:iw4 1002 1353 & 1.221 & $>67.8$ & 0.62 & $<10.0$ & 4 & 0.513 & 2.6 & 4  
   & $<0.91$\\
   CFRS 14.0547 & 1.160 & $>348$ & 0.57 & $<1.8$ & 4 & 1.55 & 2.2 & 4  
   & $<0.24$\\
   CFRS 14.0154 & 1.1583 & $>192$ & 0.57 & $<3.4$ & 4 & 0.610 & 2.9 & 4  
   & $<0.24$\\
   SSA 22-16 & 1.36 & $>165$ & 0.74 & $<4.6$ & 4 & 1.05 & 0.20 & 4 & $<3.8$\\
   CFRS 22.1153 & 1.3118 & $>125$ & 0.54 & $<5.0$ & 4 & 1.00 & 1.6 & 4  
   & $<1.1$\\
   \hline
   Nearby galaxies & & & & & & & & &\\
   Spectroscopy & & & & & & & & &\\
   Mrk 54 & 0.0448 & $>111$ & ... & $<2.7$ & 2 & 0.83 & 0.92 & 5 & $<1.2$\\
   Mrk 496 & 0.0293 & $>7.5$ & ... & $<40$ & 2 & 0.13 & 2.4 & 6 & $<4.4$\\
   Mrk 66 & 0.0218 & $>7.6$ & ... & $<39$ & 2 & 0.064 & 1.2 & 6 & $<13$\\
   Mrk 1267 & 0.0193 & $>7.8$ & ... & $<38$ & 2 & 0.068 & 3.5 & 6 & $<1.5$\\
   IRAS 08339+6517 & 0.0187 & $>12$ & ... & $<25$ & 2 & 0.34 & 1.6 & 6 
   & $<5.7$\\
   \hline
  \end{tabular}
 \end{minipage}
\end{table*}%

\section{Discussions}

\subsection{Comparison with other observations of Lyman continuum}

In the following section, we compare the limits obtained in Sect.~4.4 for our
two galaxies with those already presented in the literature (see Sect.~1). 
To do this, we have compiled all the UV-to-LC flux density ratios published in
refereed journals. The data are summarized in Table 4, where they are divided
into several categories depending on the source redshift and the observational
technique. For the broad-band $U_{300}$ data, we restrict ourselves to the
galaxies with $z_{\rm sp} > 3.33$ in order to avoid any contamination of
nonionizing UV photons. Thus, we took three galaxies from the list of
FS03.\footnote{There is one more galaxy (HDFN:[FLY99] 688) satisfying the 
redshift criterion in the list of FS03, but it is detected by $Chandra$ as 
a X-ray source \citep{hor01}. Hence, we remove it from our discussions because 
it may be an AGN.}

We first compare the observed UV-to-LC flux density ratio 
($F_{\rm UV}/F_{\rm LC}$, column 3 in Table 4) and the estimated relative
escape fraction ($f_{\rm esc,rel}$, column 5), using the intrinsic luminosity
density ratios (Sect.~4.2) and the IGM opacity model (Sect.~4.3, column 5),
which are tailored to the specific wavelength and the filter bandpass (column
6) for each individual galaxy.

\subsubsection{$z\sim3$ galaxies}

The column 3 in the top of Table 4 shows that our two observed lower limits of 
$F_{\rm UV}/F_{\rm LC}$ are comparable to those already obtained for galaxies
at $z\sim3$. 
G02 have reached constraints about two times better than ours, because their
two objects are more than two times intrinsically brighter. Our objects, as
those of S01, are comparable in luminosity to the $L_*$ of $z\sim 3$ LBGs 
\citep[see column 7; $L_*(z=3)\sim 1\times10^{29}$ erg s$^{-1}$ Hz$^{-1}$;]
[]{ste99,ade00}.

In terms of relative escape fraction (column 5), we add one more
object (HDFS 85) to the two galaxies of G02 that have been reported with a
relative escape fraction lower than the detection of S01. 
Because of the small sample size, these lower limits remain consistent with
the relative escape fraction of S01. Although we have applied a mean IGM
opacity for simplicity, a line of sight with larger than average
opacity cannot be ruled out for an individual galaxy (not to speak of an
unusual object as the escape is probably a random phenomenon).

The broad-band $U_{300}$ photometry in the HDFN reaches similar lower limits
on $F_{\rm UV}/F_{\rm LC}$ (it would have reached slightly better values at
the same luminosity level as ours). This, however, does not result in
interesting constraints on the relative escape fraction because of the IGM
opacity. The IGM opacity through the $U_{300}$ filter, with an effective
wavelength of 700 \AA\ in the source rest-frame, is much larger than those at
wavelengths closer to the rest 900 \AA\ (column 4). Since meaningful
constraints on the relative escape fraction are not reached, this sample will
be removed from discussions in terms of the absolute escape fraction presented
below. 

The top of Table 4 is also interesting for a comparison of methods of
observations and a validation of future approaches. As expected, the
narrow-band photometry can provide better constraints than spectroscopy. 
The better constraints of G02 are actually due to the high luminosity of
the objects. If our sample had a similar luminosity to the galaxies of
G02, say, $L=4\times10^{29}$ erg s$^{-1}$ Hz$^{-1}$, we would have obtained 
$F_{\rm UV}/F_{\rm LC}\ga40$ (3-$\sigma$ level). The broad-band photometry is
disqualified by the IGM opacity contribution (or requires the help of models
that lead to less direct constraints). The narrow-band photometry can
provide, even for a single $L_*$ LBG, a lower limit on $F_{\rm UV}/F_{\rm LC}$
comparable with that measured in the composite spectrum of 29 galaxies by
S01. Stacking a significant number of galaxies observed through a narrow-band
filter will be able to go much deeper. We can also average the IGM opacity
against any unusual line of sight in the stacking process. Such an
approach would allow a significant comparison with the result of S01 and
reveal which fraction of high-$z$ galaxies have high relative LC escape
fraction.

\subsubsection{$z\sim1$ galaxies}

Thanks to the high sensitivity of the HST/STIS, M03 have obtained very
good lower limits on the flux density ratio of 11 galaxies at $z\sim1$, 
$F_{\rm UV}/F_{\rm LC}>70$--350 at 3-$\sigma$ confidence level (column 3 in 
the middle of Table 4). With the correction for the average IGM opacity
described in section 4.3, these ratios are converted to upper limits on the
relative escape fraction of 2--10\% (column 5). The luminosities of this
sample are similar to $L_*$ of $z\sim3$ LBGs (column 7).

The average IGM opacity (column 4) is much smaller than those at high
redshift, but not completely negligible as assumed by M03, if we consider
the Lyman limit systems (LLSs) which dominate the IGM opacity for $z\sim1$
galaxies. The rarity of LLSs favors a statistical treatment rather than the
average opacity adopted above. Based on the number distribution function of
the IGM clouds assumed here (see Appendix), the expected number of LLSs
within the wavelength range observed by M03 (1300 \AA\ $\la \lambda_{\rm obs}
\la$ 1900 \AA) is about 0.3. Roughly speaking, for one-third of the objects of
M03, the LLSs would loosen the constraints obtained under the assumption of no
IGM opacity. In this sense, the number of galaxies in M03 is large enough
to conclude a very small relative escape fraction for their objects.

\subsubsection{Nearby galaxies}

Although the upper limits on the LC from nearby galaxies have been estimated
from their H$\alpha$ fluxes directly in terms of absolute
escape fraction \citep{lei95,hur97,deh01}, it is possible to evaluate the
observed lower limits of $F_{\rm UV}/F_{\rm LC}$ for comparison with higher $z$
objects \citep[the bottom part of Table 4; Table 1 in][]{deh01}. Except for
Mrk 54, these limits, corrected for the foreground absorption by the
Galactic HI and H$_2$ gas, are not better than those obtained at high-$z$
(column 3). The low luminosity of the sample galaxies is one of the causes 
(column 7). Nevertheless, moderate upper limits are obtained for the relative
escape fraction, $< 25$--40\% (3-$\sigma$ confidence level; column 5), because  
no correction for the IGM opacity is necessary. For Mrk 54, whose luminosity
is similar to those of our sample (column 7), the observed lower limit of
\cite{deh01} translates into an upper limit on the relative escape fraction of
3\% comparable to those for $z\sim1$ galaxies of M03 (column 5).

\subsection{Absolute escape fraction}

As discussed above and shown by equation (2), estimating the {\it absolute}
escape fraction from the {\it relative} escape fraction requires an evaluation
of the dust attenuation within each galaxy which is often difficult and
uncertain \citep[e.g.,][]{bua02,pet98,meu99}.
Although the infrared to UV flux ratio is a good estimator of the UV
attenuation \citep{bua99,gor00}, the infrared fluxes for high-$z$ galaxies are
not available yet. Here, we adopt a calibration between the nonionizing UV
slope and the UV attenuation proposed by \cite{meu99} for simplicity. 
However, we should keep in mind that the calibration depends on the type of
galaxies, starburst or not \citep{bel02,kon04}.

The UV slope of the galaxies listed in Table 4 have been searched in the
literature, or, if not available, estimated from broad-band measurements with
the assumption of a power-low spectrum 
($f_\lambda \propto \lambda^{\beta}$)\footnote{In the estimation, we 
neglected the effect of the IGM HI absorption on the broad-band photometry 
because it is small enough, although this results in an overestimation of
$\beta$.} (references in column 9 of Table 4). 
This was not possible for some galaxies of M03 because their colors in the
rest-frame $\lambda<3000$ \AA\ are not available. 
For the composite spectrum of S01, we estimated the UV attenuation from the
reported mean $E(B-V)$ via the Calzetti's attenuation law \citep{cal00}. 
The UV slope of the composite spectrum shown in Fig.1 of S01 is consistent
with the slope corresponding to the estimated attenuation.
We estimated UV slopes of our two galaxies from their broad-band colors 
although we have UV spectra of our two galaxies because the data quality is
not good.\footnote{We can measure the UV slopes in the spectra. The obtained 
slopes for the two objects are similar, $\beta=-1.9\pm0.3$, which translates 
into $A_{\rm UV}=0.7\pm0.6$ mag.}

Since the calibration gives the attenuation at the rest-frame
1600 \AA, we convert it into the attenuation at the appropriate UV wavelength
by the Calzetti's attenuation law \citep{cal00} if the UV wavelength is
different from 1600 \AA\ (column 8). We note here that the uncertainty
resulting from those on the UV slope and colors is very large; for
example, the uncertainty of about 0.05 mag in $V_{606}-I_{814}$ for our two
galaxies translates into 
$\Delta A_{\rm UV} \sim 2 \Delta \beta \sim 6 \Delta (V_{606}-I_{814}) \sim 0.3$
mag \citep[see][]{meu99}.

For galaxies at $z\sim3$, we find that 
(1) the absolute escape fraction from the detection by S01 is 
$60\pm13$\% (1-$\sigma$ observational uncertainty), (2) the absolute 
escape fraction of the two brightest LBGs observed by G02 is less than 5\% 
(3-$\sigma$), (3) the absolute escape fraction of $L_*$ LBGs observed by us is
less than 20--40\% (3-$\sigma$).

For $z\sim1$ galaxies, we find very small upper limits on the 
absolute escape fraction, typically, less than a few percent 
(3-$\sigma$). Since small upper limits were obtained even for the relative
escape fraction, the conclusion of very small absolute escape fractions 
for the observed galaxies seems robust against the uncertainty of 
estimating the dust attenuation.

For nearby galaxies, we find small upper limits on the absolute escape
fraction, less than 10\% (3-$\sigma$). These values are a factor of 0.1--1
smaller than those estimated from the comparison between the LC fluxes and the
H$\alpha$ fluxes \citep{lei95,hur97,deh01}. A cause for this discrepancy may
be the LC extinction by dust in H {\sc ii} regions; LC photons are absorbed by
dust before they ionize neutral hydrogen atoms \citep[e.g.,][]{ihk01,ino01}. 
This effect leads us to underestimate the intrinsic LC flux from the
flux of the recombination line and to overestimate the escape fraction. 
However, uncertainties of UV and H$\alpha$ attenuations are also likely to
play a role.

\section{Conclusions}

We made the first attempt to constrain the amount of 
Lyman continuum (LC) escaping from galaxies, using narrow-band photometric 
observations with the VLT/FORS, and then, reached the following 
conclusions:

\begin{enumerate}

\item
Because of an unexpected systematic effect on photometric redshifts, only
two objects with a redshift appropriate for measuring the LC radiation
are left in our sample. None of these two galaxies, HDFS 85 
($z_{\rm sp}=3.170$) and HDFS 1825 ($z_{\rm sp}=3.275$), which are 
$\sim L_*$ Lyman break galaxies (LBGs), are detected at a 3-$\sigma$
level. The resulting 3-$\sigma$ lower limits of the observed UV-to-LC
flux density (per Hz) ratio are 15.6 and 10.2 for HDFS 85 and 1825,
respectively. These limits are compatible with the detection in the composite
spectrum of 29 LBGs by \cite{ste01}. \cite{gia02} obtained slightly 
better limits than ours because of the high luminosity of their two galaxies. 

\item 
After a comparison with population synthesis models and a correction for
average IGM opacity, the observed lower limits translate into relative escape
fractions less than 0.7 and 2.2 (3-$\sigma$ confidence level) for HDFS 85
and 1825, respectively. 

\item 
In addition to the two objects observed by \cite{gia02}, 
HDFS 85 makes the third case of a relative escape
fraction smaller than that reported by \cite{ste01}. Fluctuations of the IGM
opacity from line of sight to line of sight and randomness of LC
escape from object to object may play a role in this discrepancy.

\item 
Very small values of the absolute escape fraction are estimated from
the relative escape fraction, namely less than 10\%, except for a few cases. 
The dust attenuation which is difficult to evaluate from the UV data only is a
source of uncertainty.

\item 
When the high luminosity of the galaxies observed by \cite{gia02} is accounted
for, our observations show that narrow-band photometry can reach stronger
limit than spectroscopy in terms of the relative escape fraction. This
advantage is not obtained with broad-band imaging at high-$z$ because of the
IGM opacity. Stacking a significant number of deep narrow-band images
of drop-out galaxies has, therefore, the potential to confirm or not the high
relative escape fraction reported by \cite{ste01}. In addition to increasing  
sensitivity, such a method would average the IGM opacity and the randomness of
the LC escape.
\end{enumerate}

\begin{acknowledgements}

We thank Tsutomu T. Takeuchi for a lot of valuable comments, 
Matthew A. Bershady for kindly providing us with his opacity model 
as a machine-readable form, Alberto Fern{\' a}ndez-Soto for helpful
discussions, and ESO support astronomers for their cooperation during the
phase 2 submission and observations.
AKI also thanks Hiroyuki Hirashita, Masayuki Akiyama, Hideyuki Kamaya, 
and Shu-ichiro Inutsuka for their continuous encouragements. 
In the middle of this work, AKI was invited to the Laroratoire d'Astrophysique
de Marseille and financially supported by the Observatoire Astronomique de
Marseille-Provence. AKI is also supported by the JSPS Postdoctoral
Fellowships for Research Abroad.

\end{acknowledgements}

\onecolumn 
\appendix
\section{A mean IGM opacity model}

In this appendix, we first show that the number density distribution function
of the Lyman limit systems (LLSs) reasonably agrees with an extension of the
distribution function of the Lyman $\alpha$ forest, in contrast to 
previous works. That is, the intergalactic clouds may be described by a
continuous distribution from low- to high-density through any redshift. 
Second, we present an analytic approximation of a mean opacity of the
intergalactic medium (IGM) based on the distribution function.

A mean IGM opacity at an observed wavelength $\lambda_{\rm obs}$
against a source with $z=z_{\rm S}$ can be expressed as 
\citep[e.g.,][]{mol90,mad95}
\begin{equation}
 \langle\tau^{\rm IGM}_{\lambda_{\rm obs}}(z_{\rm S})\rangle
  = \int_0^{z_{\rm S}}dz\int_{N_{\rm HI,low}}^{N_{\rm HI,up}} dN_{\rm HI}
  \frac{\partial^2{\cal N}}{\partial z \partial N_{\rm HI}}
  [1-\exp\{-\sigma_{\rm H}(\lambda_{\rm obs}/1+z)N_{\rm HI} \}]\,,
\end{equation}
where $N_{\rm HI}$ is the HI column density of a cloud in the IGM, 
$N_{\rm HI,low}$ and $N_{\rm HI,up}$ are the lower and upper limit 
of the cloud distribution, and $\sigma_{\rm H}(\lambda)$ 
is the hydrogen cross section at the wavelength $\lambda$.
Here, we assume a differential number distribution function of the IGM clouds
expressed as  
\begin{equation}
 \frac{\partial^2{\cal N}}{\partial z \partial N_{\rm HI}}
 =f(N_{\rm HI})g(z)\,,
\end{equation}
where $f(N_{\rm HI})\propto N_{\rm HI}^{-\beta}$ with the normalization
as $\int_{N_{\rm HI,low}}^{N_{\rm HI,up}}f(N_{\rm HI})dN_{\rm HI}=1$ 
and $g(z)={\cal A}(1+z)^\gamma$. We note here that the normalization 
$\cal A$ means the total number of the IGM cloud per unit redshift interval 
and is a function of $N_{\rm HI,low}$ and $N_{\rm HI,up}$.

Observationally, the power-law index $\beta$ may be regarded as 
a constant equal to about 1.5 for a wide range of column densities from  
$\log N_{\rm HI}/{\rm cm^{-2}}=12.5$ to 21.5, 
although it seems to depend on redshift and column density 
in detail \citep{sar89,pet93,kim02,per03}. For simplicity, we assume 
a single power-law index $\beta=1.5$ for all redshifts and column densities.

For the cloud number evolution with redshift, \cite{mad95} 
adopted two different evolutionary indices ($\gamma$) 
for the Lyman $\alpha$ forest ($N_{\rm HI}\la10^{17}$ cm$^{-2}$) 
and for the LLSs ($N_{\rm HI}\ga10^{17}$ cm$^{-2}$), 
based on observations by \cite{mur86} and \cite{sar89}.
However, the same index $\gamma$ for these populations is compatible 
with recent observations as shown in Fig.~A.1.
The LLS number evolution is reproduced from 
\cite{per03} in this figure as the filled square points with error bars.
The solid line is the number evolution of the Lyman $\alpha$ forest  
reported by \cite{kim02} and \cite{wey98} multiplied by the reduction 
factor due to the lowest column density considered; 
\cite{kim02} deal with Lyman $\alpha$ forest with 
$\log N_{\rm HI}/{\rm cm^{-2}}=13.644$--17, whereas \cite{per03} deal with  
the clouds with a column density larger than $1.6\times10^{17}$ cm$^{-2}$, 
so that the multiplicative factor is 
$(4.4\times10^{13}/1.6\times10^{17})^{\beta-1}\approx0.017$ 
when $\beta=1.5$. There is a systematic disagreement along the vertical axis 
for high-$z$ points between the solid line and the data, but the evolutionary 
slopes are consistent as noted by \cite{per03}. The high-$z$ data 
are reproduced by the dashed line whose reduction factor is 0.01. 
On the other hand, the LLS number evolution reported by 
\cite{sar89} and adopted by \cite{mad95} is shown as the dotted line in 
Fig.~A1 and cannot reproduce the data points of \cite{per03}. 
Although the number evolution suggested by \cite{ste95} shows a better fit
(dash-dotted line), in this paper, we assume the case of the solid line. 
This means that we assume the same redshift evolution of the cloud number
density over the entire column density range with $\beta=1.5$.

\begin{figure}
 \centering
 \includegraphics[width=8cm]{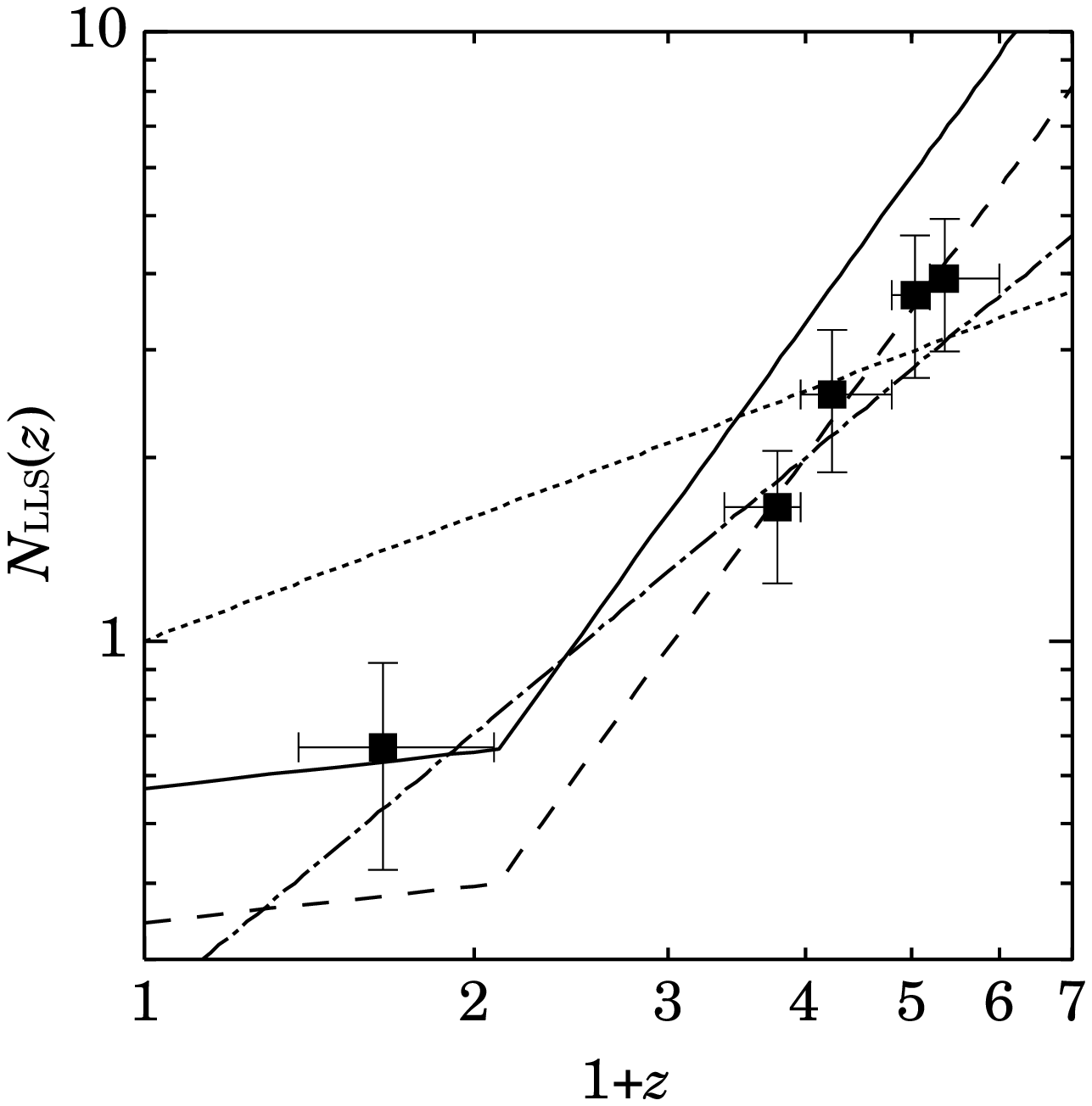}
 \caption{Number density evolution of the Lyman limit systems (LLSs; 
   $\log N_{\rm HI}/{\rm cm^{-2}}=17.2$--20) in a unit redshift interval along
   a line of sight. Data points with error-bars are
   taken from \cite{per03}. The solid line is the number evolution expected
   from that of the Lyman $\alpha$ forest  
   ($\log N_{\rm HI}/{\rm cm^{-2}}=13.64$--17) reported by \cite{kim02} and 
   \cite{wey98} with a single power-law distribution of the column density; 
   the number of LLSs is that of Lyman $\alpha$ forest multiplied by
   $(4.4\times10^{13}/1.6\times10^{17})^{\beta-1}\approx0.017$ 
   when $\beta=1.5$. The dashed line is the same as the solid line but the
   multiple factor is 0.01. The dotted line is the number evolution of LLSs
   reported by \cite{sar89}, and the dash-dotted line is that reported by
   \cite{ste95}.}
\end{figure}

Under a cloud distribution function with a single power-law index  
$\beta$ ($1<\beta<2$) and the same evolutionary index $\gamma$ 
for all range of the column density, 
we take a set of limiting column densities as 
$\sigma_{\rm H}N_{\rm HI,low}\ll1$ and $\sigma_{\rm H}N_{\rm HI,up}\gg1$, 
for example, $N_{\rm HI,low}=10^{12}$ cm$^{-2}$ and $N_{\rm HI,up}=10^{22}$ 
cm$^{-2}$. In this case, equation (A.1) can be approximated as 
\begin{equation}
 \langle\tau^{\rm IGM}_{\lambda_{\rm obs}}(z_{\rm S})\rangle \approx 
 \int_0^{z_{\rm S}} 
 \Gamma(2-\beta) {\cal A} N_{\rm HI,low}^{\beta-1} (1+z)^\gamma 
  \sigma_{\rm H}(\lambda_{\rm obs}/1+z)^{\beta-1}dz\,,
\end{equation}
where $\Gamma(2-\beta)$ is the usual Gamma function \citep{zuo93}. 
The factor, ${\cal A} N_{\rm HI,low}^{\beta-1}$ can be estimated from 
the observed number of clouds for a limited range of the column density 
with the power-law distribution. 
If the number of clouds with a column density between $N_{\rm l}$ and 
$N_{\rm u}$ ($N_{\rm HI,low}\leq N_{\rm l}< N_{\rm u}\leq N_{\rm HI,up}$)
is denoted as $\cal A'$, we have 
${\cal A}N_{\rm HI,low}^{\beta-1}\approx{\cal A}'N_{\rm l}^{\beta-1}$
when $N_{\rm HI,up}/N_{\rm HI,low}\gg1$, $N_{\rm u}/N_{\rm HI,low}\gg1$, 
and $\beta>1$. According to \cite{kim02} and \cite{wey98}, we adopt 
$({\cal A}',\gamma)=(6,2.5)$ for $z>1.1$ and (34,0.2) for $z\leq1.1$
against $\log N_{\rm HI}/{\rm cm^{-2}}=13.64$--17 (i.e. 
$N_{\rm l}=4.4\times10^{13}$ cm$^{-2}$).

We will present an analytic approximation of equation (A.3), 
although we can integrate it numerically with a detailed function 
of the hydrogen cross section, $\sigma_{\rm H}$. We adopt an approximated form
of the cross section for the analytical formula. 
The point is that the cross section of the $i$-th line, $\sigma_i(\lambda)$, 
can be neglected for a wavelength $\lambda$ out of a small range of 
$|\lambda-\lambda_i| \la b/c$, where $\lambda_i$ is the central 
wavelength of the line, $b$ is the Doppler parameter, and $c$ is 
the speed of light. Besides, the cross section for LC, 
$\sigma_{\rm LC}(\lambda)$, is zero when $\lambda>\lambda_{\rm L}=912$\AA. 
Thus, we can approximate $\sigma_{\rm H}(\lambda)^{\beta-1}$ 
($=[\sigma_{\rm LC}(\lambda)+\sum_i \sigma_i(\lambda)]^{\beta-1}$) 
to $\sigma_{\rm LC}(\lambda)^{\beta-1}+\sum_i \sigma_i(\lambda)^{\beta-1}$, 
because only one term has meaningful value for a specific wavelength $\lambda$.

Adopting this approximation, we can reduce equation (A.3) to 
\begin{equation}
 \langle\tau^{\rm IGM}_{\lambda_{\rm obs}}(z_{\rm S})\rangle \approx 
 \tau_{\lambda_{\rm obs}}^{\rm LC}
 + \sum_i \tau_{\lambda_{\rm obs}}^i\,,
\end{equation}
where
\begin{equation}
 \tau_{\lambda_{\rm obs}}^{\rm LC}= \cases{
   \Gamma(2-\beta) (\sigma_{\rm L} N_{\rm HI,low})^{\beta-1} 
   \left(\frac{\lambda_{\rm obs}}{\lambda_{\rm L}}\right)^{3(\beta-1)}
   \int_{z_{\rm min}}^{z_{\rm S}} {\cal A} (1+z)^{\gamma-3\beta+3} dz
   & (for $\lambda_{\rm obs}<\lambda_{\rm L}(1+z_{\rm S})$ ) \cr
   0 & (otherwise) \cr } \,,
\end{equation}
and
\begin{equation}
 \tau_{\lambda_{\rm obs}}^i= \cases{
   \Gamma(2-\beta) (\sigma_{0,i} N_{\rm HI,low})^{\beta-1}  {\cal A}
   \left(\frac{\delta b}{c}\right) 
   \left(\frac{\lambda_{\rm obs}}{\lambda_i}\right)^{\gamma+1}
   & (for $\lambda_i < \lambda_{\rm obs} < \lambda_i (1+z_{\rm S})$) \cr
   0 & (otherwise) \cr }\,.
\end{equation}
For the Lyman continuum, 
$\sigma_{\rm H}(\lambda)=\sigma_{\rm L}(\lambda/\lambda_{\rm L})^3$ 
with $\sigma_{\rm L}=6.30\times10^{-18}$ cm$^2$ \citep{ost89} 
being the cross section at the Lyman limit. 
The integral in the last term of equation (A.5) can be done
analytically if $z$ dependences of $\gamma$ and $\cal A$ are simple.
The lower edge of the integration is  
$z_{\rm min}=\max[0,(\lambda_{\rm obs}/\lambda_{\rm L})-1]$.
For Lyman series lines, whose central wavelength and cross section are 
denoted as $\lambda_i$ and $\sigma_{0,i}$, 
we have integrated the integral in equation (A.3) over a narrow range 
characterized by the Doppler parameter, $b$, around the line center, 
assuming a rectangle line profile. Actually, the integration has been made 
over the range of the Doppler width multiplied by a numerical factor 
$\delta$ which is order of unity. From the comparison between a numerical 
integration of equation (A.3) with a detailed function of hydrogen cross
section\footnote{We approximately take into account the Voigt profile 
by the following equation; $\phi(x)\approx \exp(-x^2)+a/\sqrt\pi x^2$, 
where $a$ is the damping constant, and $x$ is the relative displacement 
in the frequency space, i.e. $x=(\nu-\nu_i) / \Delta \nu_{\rm D}$, where 
$\nu$, $\nu_i$, and $\Delta \nu_{\rm D}$ are the frequency, the central 
frequency of the line, and the Doppler width in the frequency space, 
respectively \citep{fer96}.}
and the analytic approximation of equation (A.4--6), 
we find that the approximation becomes excellent 
if $\delta=2.6$ for $b=30$ km s$^{-1}$, which is a typical Doppler parameter 
for Lyman $\alpha$ forest \citep{kim02}.
All calculations include up to 40-th line of Lyman series whose central 
cross sections are calculated by the data taken from \cite{wie66}.

Fig.~A.2 shows the mean IGM opacities for several source redshifts
calculated by equations (A.4--6) with the distribution function based on 
\cite{kim02} and $(\delta,b)=(2.6,30\,{\rm km\,s}^{-1})$. 
Since Kim's observations are made against 
Lyman $\alpha$ forest below $z\sim4$, we have shown cases up to 
$z_{\rm S}=4$. As a comparison with a previous opacity model, we also plot 
the result by \cite{ber99} (the case of ``MC-Kim'') as the dotted curve 
in the figure. Since they adopted the LLS number evolution 
of \cite{sar89} as done by \cite{mad95}, there are some discrepancies 
between ours and theirs; our larger (smaller) optical depth at around 
the observed 3000 (1500) \AA\ than \cite{ber99} is caused by our larger 
(smaller) number of LLSs at $z\ga2$ ($\la2$) than \cite{sar89} 
as shown in Fig.~A.1. However, the quantitative agreement is reasonably 
good.

As another test, we compare our optical depth with that measured by 
\cite{ste01}. From a comparison between two composite spectra of galaxies 
and QSOs at $\langle z\rangle \simeq 3.4$ and 3.3, respectively, 
\cite{ste01} have found a mean IGM optical depth of 1.35 for the rest-frame 
900 \AA, whereas our model gives 1.36 for the wavelength and $z=3.4$ source. 
We have found an excellent agreement with each other.

\begin{figure}
 \centering
 \includegraphics[width=9cm]{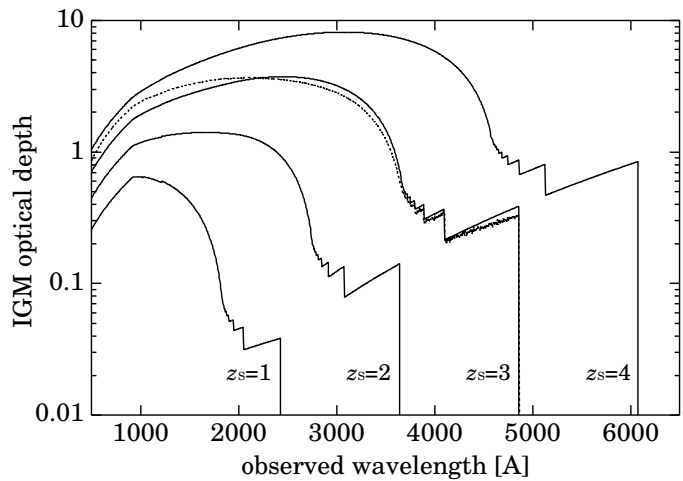}
 \caption{Examples of the IGM optical depth calculated by an analytic
   approximation presented as equations (A.4--6). The dotted curve is 
   the model ``MC-Kim'' by \cite{ber99}.}
\end{figure}

\end{document}